SNOWMASS Instrumentation and Calorimetry
*High Rate and High Precision Timing and Calorimeter Detectors*

David R. Winn[1*], Y. Onel[2], B. Bilki[2]

[1]Fairfield University, [2]University of Iowa

*winn@fairfield.edu
ORCID 0000-0003-2637-5743

*Introduction:* High precision timing, high rate calorimeters, and radiation resistance are becoming an important issue in particle physics especially in Energy and Intensity Frontiers. We discuss doped Zinc Oxide (ZnO:Ga or GZO; ZnO:X where X is Al, Cu or others) as a very fast scintillator and wavelength shifter (WLS), Total internal reflection films, and PMT capable of counting at 300 MHz with 10 ps timing precision, with superior radiation resistance.

• *Rate/Occupancy/Hysteresis:* In colliders, present very forward calorimeter or timing channels should operate at 40 MHz with some proposed colliders or upgrades up to and perhaps exceeding 100 MHz, with calorimeter hysteresis pulse-pulse <2%. Signals with FW10%-10%Max <10ns and segmentation to handle >200 pileup (PU) are advantageous in some future colliders and upgrades, where the bunch crossing times are proposed to be lowered to 5-10ns. The *energy scale* is ~0.1 GeV (1/4 minimum ionizing particles - MIPs) to ~1 TeV per calorimeter channel, a dynamic range of ~$10^4$ so crossing-crossing hysteresis is an issue. Similar to proposed new colliders, tagged neutrino beams (from pions or muon factories) and tagged kaon beams would benefit from rep-rates exceeding 100's of MHz. The high track density and pile-up in high luminosity particle colliders are challenges for event reconstruction and analysis. MIP (minimum ionizing particle) pileup is a few percent in ~1x1 cm, 1200 cm radially along η=0. Occupancy/Rate: At ~10 m from the interaction in the η≥3 forward regions at HLHC/SLHC, the occupancy of a 1 $cm^2$ patch of detector (calorimeter, MIP timing detectors, pre-radiators) is 100% per crossing, with pileups of ~150-500 events/crossing at highest luminosities.

• *Timing:* The case for adding a timing 4 dimension to calorimetry and tracking is becoming compelling. For pileup mitigation (100's of interactions per crossing), a time precision for mips must be ~30ps or better, with energy flow capability, to add the "4[th]-dimension" to detector systems.[1] Timing detectors must withstand 50 MRad and neutrons >$3x10^{15}$ n/$cm^2$. Timing better than ±30ps has been shown by CMS and ATLAS to improve $E_T$miss resolution, and tag secondary vertices to ±few mm. Precise timing of calorimeter deposits and vertexes enable rejection of spurious data inconsistent with the primary vertex time. We discuss fast scintillators and WLS for MIPs capable of timing precision to ±10's ps, and rate capabilities exceeding 100's of MHz. Issues for defining a Figure of Merit for timing scales as $\tau$decay/$\sqrt{N}$electrons, and the FOM for rate capability scales inversely as $\tau_{decay}$.

• *Radiation Damage:* Calorimeters at near and far future pp, ep, e+e-, e-ion, and muon colliders will need to measure high-energy depositions at irradiation levels approaching/exceeding GigaRads and >$10^{17}$ neutrons/$cm^2$ in the forward regions. In planned LHC energy and luminosity upgrades over 20 years, at 3,000-5,000 $fb^{-1}$, the 1 MeV equivalent neutron flux is ~$10^8$/$cm^2$/s at η≥4.9 and at ~10 m from the IP [2] [3]. At η≥ 4.9, a calorimeter 10m from the IP has a dose ≥ GigaRads in the 1[st] ~2λ.

*Physics:* At hadron-hadron, lepton-hadron, and lepton-lepton colliders, physics processes include vector boson fusion and scattering where color is not exchanged; Higgsstrahlung; triple Higgs coupling H−>HH; Higgs to invisible H−>XX(Dark Matter); Higgs decays H−>ττ (μμ); luminosity monitors. S/N requires jets at η>2.8 (η=3.1), and a rapidity difference between tagging jets $\Delta\eta_{ij}$ > 5 for cleanest S/N. The 125 GeV Higgs at higher √s enhances Higgs decay activity boosted more into the forward region. High rates, Time-of-flight and radiation resistance are part of many fixed target experiments and factory-collider experiments. Examples include: LHCb, Belle/BES-III or future factory calorimeter, TOF, polarimeter and luminosity systems. Present and future rare K decay and some dark photon (beam-dump) experiments require high rate calorimetry. In the future, tagged neutrino beams to obtain purer beams of electron or muon neutrinos [4], and tagged kaon beams[5] are proposed, and counting rates in particle telescopes greatly exceeding >100 MHz counting rates are highly desired. Muon lepton violation experiments also benefit from very high counting rates. At the present time, the basic sensors used in LHC forward calorimetry or for high precision timing (collected ions/electrons e.g. Si, LArgon, gasses; optical signals from scintillator or Cerenkov radiators as detected by SiPMs) can *neither* survive the raddam *nor* operate sufficiently fast – rates >100 MHz or Timing precision <30ps. We propose scintillators and WLS fibers with ~few ns decay times and radiation resistance for systems that must operate at minimum for 10 years with few failures or degradation in performance.

**Fast Scintillator and Wavelength Shifter:**
GZO: We propose ZnO:Ga (GZO) and ZnO:X (X=Cu, Al and others) as a fast and highly radiation resistant scintillator, WLS (wavelength shifter) and similar low-index total internal reflection (TIR) cladding films, with rate capability >100MHz, time resolution <30 ps, able to survive ~GigaRad radiation doses. For optical WLS or Scintillating fibers, thin film deposition techniques including Atomic Layer Deposition (ALD)[6] of Zn:X on quartz fiber armatures are possible and presently used commercially as transparent conducting electrodes in displays and solar cells.

Optical signals with >100 MHz rates require scintillator decay constants <1ns, or Cherenkov radiators. Scintillators with high figure of merit include ZnO:Ga(GZO) (0.6-0.7 ns decay) and LYSO. GZO is favorable for timing and for radiation resistance – zinc oxide is a metal oxide used as a rad-hard CRT phosphor, much more radiation resistant and faster than LYSO. However, GZO is difficult to grow as single crystals thicker than a few mm (see Fig 1). Deposition by MOCVD, PVD (e-beam), and/or ALD are preferred for use as a film scintillator or WLS on quartz (50µm quartz 10 µm ZnO:X.

| **Fast Scintillators** | | |
|---|---|---|
| | ZnO:Ga (GZO) | LYSO |
| Light: γ/MeV | 12,000-15,000 | 40,000 |
| Decay (ns) | 0.6-0.7 | 41-43 |
| $t_{rise}$ | 42 ps | 70 ps |
| γ/MeV/ns | 6,000 | 740 |
| $T_{melt}$ (°C) | 1,980 | 2,050 |
| dE/dx: MeV/cm | 8.4 | 9.55 |
| Xo (cm) | 2.51 | 1.14 |
| Peak λ (nm) | 390-375 | 420 |
| Index n | 1.85 | 1.82 |
| Density (g/cc) | 5.6 | 7.4 |
| $\sqrt{T_{rise}} \times \sqrt{T_{fall}}$ (ps) | 167 | 1700 |
| | | |

The photographs below show the largest thicknesses able to be made at present from melt-techniques; even at ~3mm thickness the crystallite boundaries affect the optical clarity. ALD can in principle be scaled up to make thicker and more transparent by depositing on thin (~50µm thick 30 cm wafers are routinely handled with air-chucks in foundries) tiles of quartz, synthetic sapphire($Al_2O_3$), some radiation-resistant glass compositions, and newly available clear colorless polyimides(Kapton), and then stacking the resulting tiles. ALD tooling for films on 75" displays are in use.

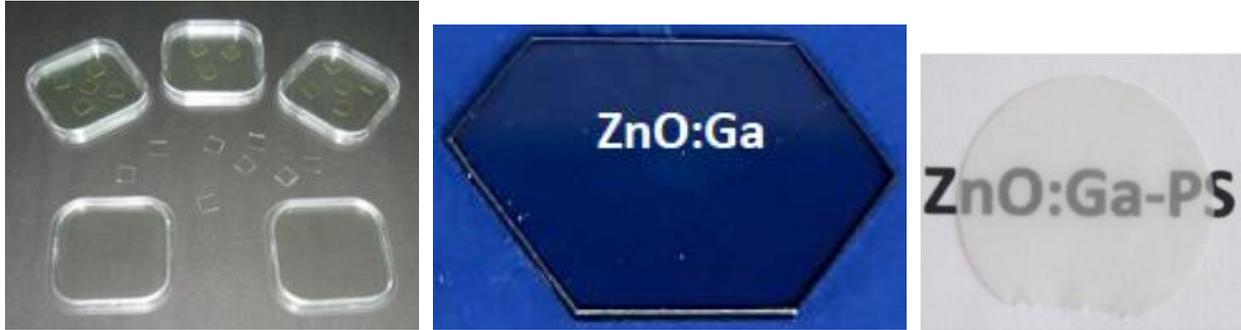

*Fig.1:* L: 1cmx1cmx0.9mm ZnO wafers (Semiconductor Wafer Inc.) (ignore the plastic boxes)M: ~3x3cm x 1.9 mm ZnO:Ga tile -FJIRSM(China) – microcrystallinity affects the optical transmission.R: 2"x0.5mm Polystyrene disc loaded with 10% ZnO:Ga nanopowder[7] Note the small thicknesses.

**Direct comparisons of LYSO and GZO pulses:**

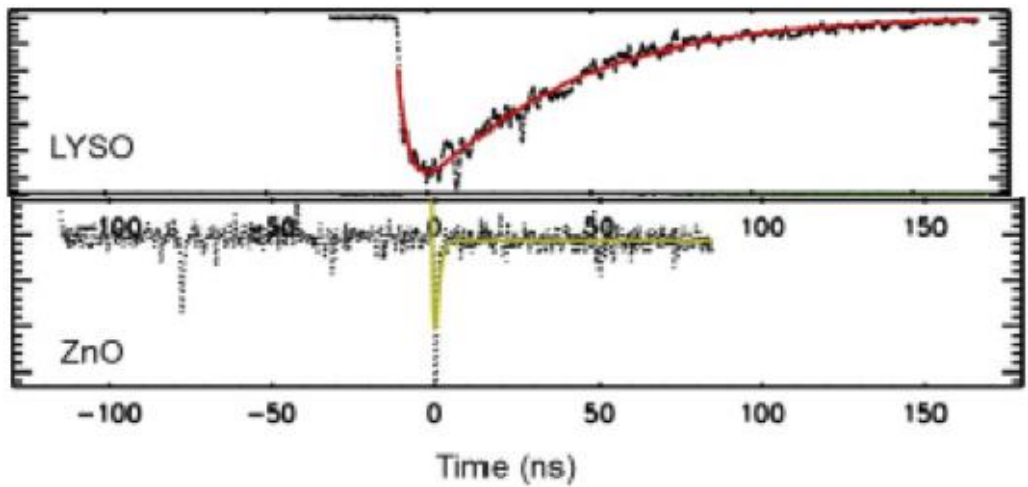

*Fig.2:* Comparison of LYSO(red) and ZnO:Ga(yellow) flash-X-ray pulse shapes using a 20 GHz bandwidth 100 GHz sampling oscilloscope (y-axis arbitrary scale) using a fast MCP-PMT [8].

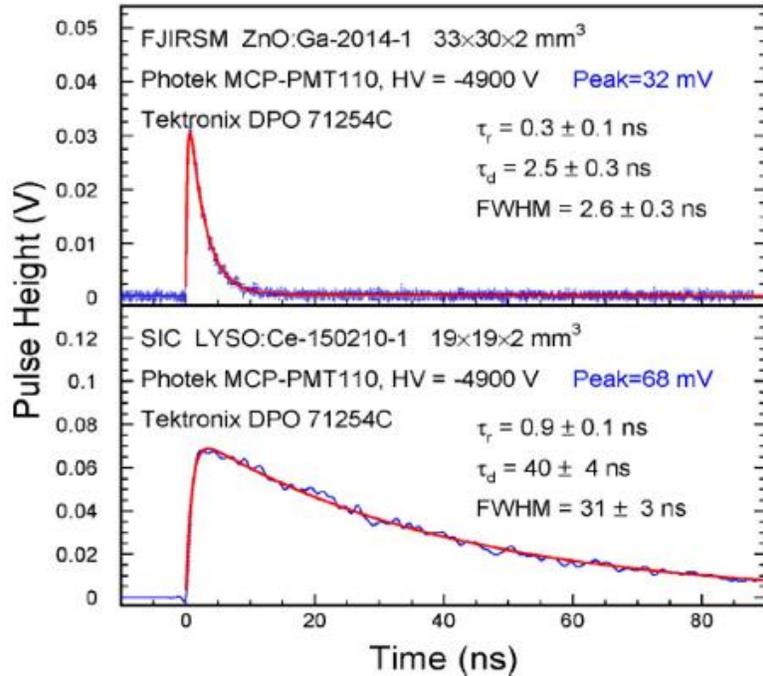

*Fig 3:* Flash x-ray stimulated ZnO:Ga emission(top) compared to LYSO using an MCP-PMT and oscilloscope through 10 m cables.[8]

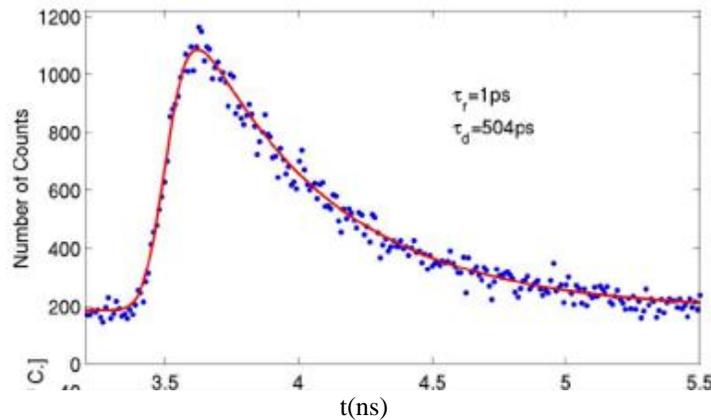

*Fig.4:* Pulse shape from a ~1 mm thick polystyrene disc loaded with 10% ZnO:Ga nanopowder (see pictures below) with a 504 ps fitted decay[9]. 2.5 ns integrates >90% of charge.

- ***ZnO:Ga (GZO) Scintillator****:* ZnO:Ga is a promising alternative scintillator (compare to LYSO – See Table 1 and Figures 2-4 above).

    1. ***Rise and Fall Time:*** ZnO:Ga(1%-5%) has a rise-time 30-40ps and a decay-time 0.5-0.7ns with a $\sqrt{T_{rise}} \times \sqrt{T_{fall}}$ product $\leq 167$ps, *10 times less than LYSO*. For 30 ps timing, > 32 p.e. are required.

    2. ***Large photon yield per ns per MeV*:** highest of any known scintillator - ZnO:Ga more visible 7,000-9,000 photons/MeV/ns, at 375-395nm with no glow.

    3. ***Pileup*** is absent for ZnO:Ga even at 300 MHz of charged particles/$cm^2$, with 90% integration <3ns.

4. *Noise***: <3ns integration reduces SiPM or PMT noise+cooling requirements, implying Rates > 100MHz

    5. *Radiation length* 2.51cm. For MIP-detecting tiles used for timing systems, relatively few background low energy gamma and x-rays will convert in GZO tiles, lowering background and noise rates compared to other dense scintillators.

    6. *Energy loss* and an *index n* are 0.84 MeV/mm and 1.85. A 500µm thick layer of ZnO:Ga produces >600 photons over 500 ps.

    7. **Doped-ZnO is very rad-hard** (as are most metal oxides) –a fast phosphor in e-beam/CRT and used in fluorescent lights. As a CRT phosphor it has >1GRad resistance (25 KeV electrons) (compare to ~10 MRad for LYSO).

- *Pileup* caused by *all radiation* (not just MIPs: x-rays, gamma, n, few Mev electrons) at an LHC IP is largely absent for ZnO:Ga compared to LYSO as used as timing detectors in front of calorimeters. A 90% integration time is <3ns (lowering SiPM noise and reducing cooling requirements) with rate capabilities shown to exceed 100 MHz counting rates[10],[11] is possible.

- *The radiation length* of GZO (ZnO:Ga) is 2.2 times *larger* than LYSO. For MIP-detecting timing tiles, this means that fewer low energy gamma and x-rays will convert in the tiles, lowering background and noise rates. LYSO is very sensitive to ~few MeV depositions, thereby simulating MIPs – an important consideration.

Comments:
1. The pulse heights LYSO vs ZnO:Ga depend on the QE(λ) of the MCP-PMT; for ZnO:Ga best performance the SiPM has to be sufficiently sensitive in ~375-390nm

2. As with LYSO, the optimal properties depend on the quality of the fabrication, and with ZnO:Ga, also optimizing the Ga doping.

3. ZnO:Ga is intrinsically cheaper than LYSO.

4. Quartz melts 260°C lower than ZnO:Ga, and so nanopowders could be loaded into rad-hard quartz like the disc shown in the figures. Similarly, silicone elastomers (ex: Sylgard 184) and the newly available clear and colorless kaptons are 100 MRad radiation resistant.

**Commercialization:** ZnO:X is now a highly commercial product and subject to considerable research as a transparent conducting electrode for solar cells and more rugged replacement for ITO which is expensive and easily damaged. ZnO:X films are now made in very large areas.

- *ZnO:Al (AZO) Wavelength Shifter(WLS):* AZO area highly efficient WLS that shifts from ~375nm peak absorption to ~500 nm green emission. Films can be formed on top of near UV scintillators or directly for quartz armatures to make WLS fibers. A total internal reflection film must be overdeposited.

- *$MgF_2$:* Magnesium fluoride films can also be formed by CVD, PVD and atomic layer deposition (ALD) processes and are compatible with deposition on GZO or AZO. The refractive indices 1.34–1.42 are ideal to form optical total internal reflection (TIR) claddings on tiles or fibers

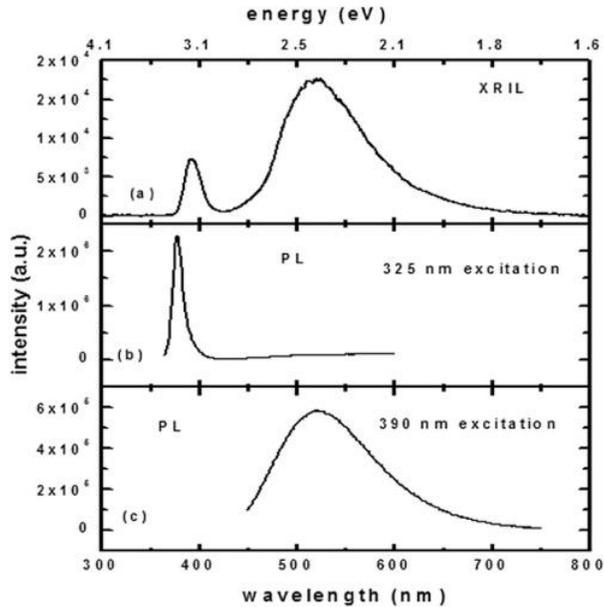

**Figure 5:** ZnO:Al (AZO) wavelength shifting – the bottom figure blue-violet shifting to green.

We suggest ALD to coat large area thin transparent(quartz) tiles coated with ZnO:Ga (GZO) scintillator films, quartz optical fibers coated with ZnO:Al (AZO) UV/blue to green wavelength shifter(WLS) films, and MgF2 low index cladding films for both, in order to make large areas and high quality crystalline material with the highest light outputs. ALD is a promising technology now being applied in commercial fab-tooling over ~2m$^2$ areas for displays, and doped ZnO films are commercially used for transparent electrodes.

Fast PMT for timing and rate: Relatively new PMT from Hamamatsu exhibit 0.4 ns risetime, 0.6 ns FWHM for 1 p.e., with 5% of max gain in 1T .

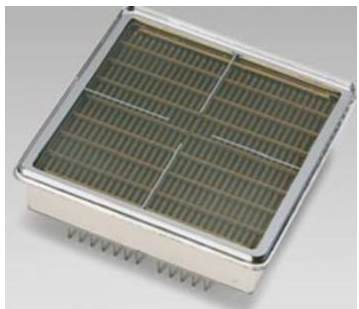 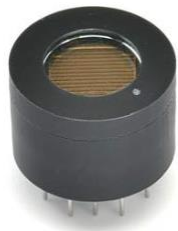 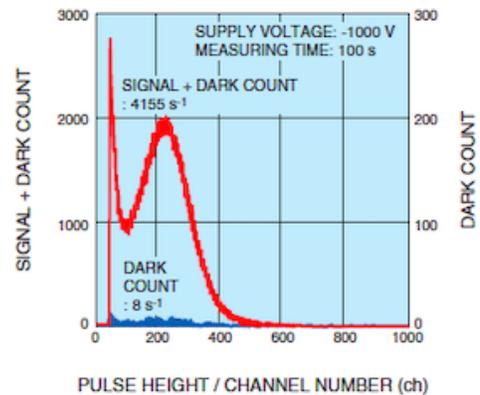

Figures 6: Left to right: MAPMT Metal envelope, B-operable 5x5cm x1.4cm thick 64 anode MAPMT(H12700B); PMT 9mm R14755U; Superior single p.e. resolution (QE$_{peak}$ ~35%).

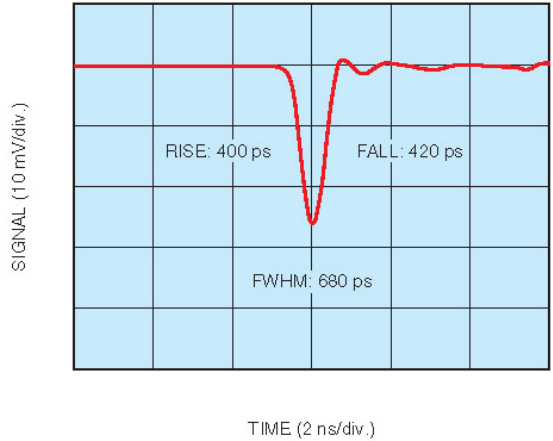

*Timing:* $T_{rise}, T_{fall}$ ~0.4ns -*FWHM of the 1 p.e. pulse < 700ps.* Hamamatsu demonstrated counting at 800 MHz

**Radiation Resistance of PMT:**
Comparative tests on gamma- and neutron- irradiation of industrial photo-multipliers and various PMT window materials confirm that permanent degradation of PMT sensitivity depends almost entirely on PMT window transmission loss [12], and not on photocathode or dynode degradation. Exposures of standard glass PMT to 1- and 2-MeV electrons show large increases in dark current *during the exposure* which are *transient*, shown to be almost entirely due to induced fluorescence in soda-lime, borosilicate or UV glass. Tubes with sapphire or quartz windows but glass tube bodies had substantially lower dark current rise during flood exposure to electrons but returned to normal dark current after the exposure ended [13]. The ATLAS experiment exposed 2 different quartz window-but glass body bialkali PMT to ~3 x $10^{14}$ n/cm$^2$ [14]. The dark current just after exposure increased by a factor of 10 in a 13 mm diameter PMT and by a factor of 7 in a 10 mm diameter PMT, both of which recovered to ~90% in 10 weeks post irradiation. However there was no obvious effect either on the QE of the bialkali photocathode or the gain. Induced light glow from the tube body glass was consistent with the dark current.

*Residual Gas:* A caution is that PMT with residual gas can damage the photocathode by reverse flowing positive ions formed by the dynode electron cloud; these ions directly impinging on the photocathode both cause after-pulses and can damage the few 10's of atomic layers thick photocathode. Studies are needed for photocathode damage under high exposures using commercial metal envelope PMT with quartz windows.

Questions to be answered:
- What is the scintillation light yield from thin GZO films for electron mips?
- What is the risetime and decay time of thin GZO scintillation?
- What is the WLS efficiency of thin AZO films for 390 nm light shifted to ~500 nm
- What is the absorption depth for 390 nm light for thin AZO films
- What is the ALD GZO and AZO film thickness growth rate?
- What is the crystallinity of the ALD ZnO based films?

*Atomic Layer Deposition of ZnO:X.* ALD (atomic layer deposition)[15] and atomic layer PVD (physical vapor deposition) is a method of fabrication that is becoming standard in VLSI fabs. Tools that work with 30cm Si wafers are now standard, and tools to coat very large areas (~m$^2$) are in use and being further developed for advanced image displays(TV/video). ALD is the best method to create pure conformal films. An alternate method by CVD (chemical vapor deposition) nearly as good and is faster, but does not create as pristine, conformal or crystalline a material. The rapidly escalating cost of indium tin oxide has led to search for lower cost transparent conductor options such as heavily doped zinc oxide. ZnO is an important technological material, which can be doped to modulate structure and composition to tailor a wide variety of optical and electronic properties[16]. ALD doped ZnO is viewed as a transparent conducting oxide for application in solar cells, flexible transparent electronics, and light-emitting diodes. To date, there are 22 elements that have been reported as dopants in ZnO via ALD. Gallium doped ZnO (GZO), aluminum (AZO) and copper (CZO) with ozone as the oxidant are readily formed at temperatures ~250-300C° [17] [18]. Metal-containing precursors include diethyl zinc, trimethyl aluminum, triethylgallium and trimethylcopper. Homogeneous films with about 2% aluminum or gallium have resistivities ~0.5 milliOhm-cm.

*GZO, AZO:* At ~200 °C, a growth thickness rate of 0.33 Å/cycle is possible. As of 2020, cycles can take as few as 0.3-1 seconds. We conservatively anticipate ~0.5-1 minute per nm, about 5-10 hour per 1µm, realistically 2 µm per 24 hour day. Five days of growth would be sufficient for a 10 µm thick ALD conformal coating of a ZnO:Al WLS film or a ZNO:Ga scintillating film on a bare quartz optical fiber. A quartz optical fiber between 100-300µm diameter would be obtained by stripping the plastic cladding from a quartz fiber. A ~30 µm scintillating film on a tile could be made in about 2 weeks of 90% continuous operation at an estimated $100/hour. About 10-20 µm is enough to demonstrate whether these are interesting materials by measuring scintillation with radiosources, and photoluminescence with lasers and LED for luminescent efficiencies.

*$MgF_2$:* Magnesium fluoride is an ultraviolet (UV) transparent material which is widely used in optical applications over a wide wavelength range. An ALD process for depositing $MgF_2$ films at 250–400 °C uses $Mg(thd)_2$ and $TiF_4$ as precursors. The growth rate for polycrystalline films (index ~1.38) is ~1.5-2 Å $cycle^{-1}$ at 250 °C. A ~3-4 µm thick film for cladding the GZO, AZO coated fibers and tiles would require about 2 days of growth. Uncoated bare quartz optical quality fibers will also be coated with ALD magnesium fluoride as a comparison.